\documentclass[a4paper,11pt]{article}
\usepackage{jinstpub} 
\usepackage{lineno}
\usepackage{float}
\usepackage{here}
\usepackage{subfigure}

\usepackage{soul,color,xcolor}

\title{\boldmath MuonSLab: A plastic scintillator based detector for muon measurement in the deep ocean }

\author[a,1]{Jiacheng Wu,}
\author[b,1]{Weilun Huang,\note{These two authors contributed equally to this work.}}
\author[b]{Ruike Cao,}
\author[b]{Qichao Chang,}
\author[c]{Wang Ding,}
\author[b]{Jingtao Huang,}
\author[d]{Liang Li,}
\author[b]{Xinchen Li,}
\author[b,d,e,2]{Hualin Mei,\note{Corresponding author.}}
\author[d]{Cen Mo,}
\author[b]{Hengbin Shao,}
\author[b]{Wei Tian,}
\author[e]{Xinliang Tian,}
\author[f]{Yichen Tian,}
\author[b,d,e]{Xin Xiang,}
\author[b,d,e]{Donglian Xu,}
\author[b]{Fuyudi Zhang,}
\author[d]{Wei Zhi,}
\author[d]{Yiwei Zhu}

\affiliation[a]{School of Electronics, Information and Electrical Engineering, Shanghai Jiao Tong University,\\
800 Dongchuan Rd, Minhang District, Shanghai, China}
\affiliation[b]{State Key Laboratory of Dark Matter Physics, Tsung-Dao Lee Institute, Shanghai Jiao Tong University,\\
1 Lisuo Road, Pudong New Area, Shanghai, China}
\affiliation[c]{Guangzhou Marine Geological Survey,\\
1133 Haibin Rd, Nansha District, Guangzhou, China}
\affiliation[d]{School of Physics and Astronomy, Shanghai Jiao Tong University,\\
800 Dongchuan Rd, Minhang District, Shanghai, China}
\affiliation[e]{Hainan Research Institute, Shanghai Jiao Tong University, Hainan, China}
\affiliation[f]{School of Ocean and Civil Engineering, Shanghai Jiao Tong University,\\
800 Dongchuan Rd, Minhang District, Shanghai, China}
\affiliation[g]{School of Mechanical Engineering, Shanghai Jiao Tong University,\\
800 Dongchuan Rd, Minhang District, Shanghai, China}

\emailAdd{mei.hualin@sjtu.edu.cn}

\abstract{
Atmospheric muons are important probes for studying primary cosmic rays and extensive air showers. Additionally, they constitute a significant background for many underground and deep-sea neutrino experiments, such as TRopIcal DEep-sea Neutrino Telescope (TRIDENT). Understanding the muon flux at various depths in the deep sea is essential for validating TRIDENT simulations and guiding the development of optimized trigger strategies. This paper introduces a novel device based on plastic scintillators and silicon photomultipliers (SiPMs) named MuonSLab, which is designed to measure muon flux in the deep sea and has the potential to be extended to other atmospheric muon property measurements. We discuss the design and instrumentation of MuonSLab and present results from several muon flux measurements, demonstrating its sensitivity to muon detection and its stability during operations across multiple locations.
}


\begin{document}
\maketitle
\flushbottom

\section{Introduction}
\label{sec:intro}
Atmospheric muons are charged particles produced in cosmic ray (CR) air showers. These particles are the most abundant charged particles reaching sea level and underground due to their low energy loss in the atmosphere, making them essential probes for cosmic ray studies. The number of muons in CR air showers is related to the CR mass composition \cite{heitlermodel}. Thus, experiments have been conducted to measure the muon number, revealing a deficit in simulations compared with measurements \cite{pao_muon}, suggesting gaps in our understanding of hadronic interactions at ultra-high energies. Furthermore, these particles can reach extremely high energies, allowing them to traverse several kilometers of seawater or ice, thereby constituting a significant background for neutrino telescopes. Large-scale neutrino telescopes like IceCube \cite{IceCube} and KM3NeT \cite{KM3Net} have reported measurements of atmospheric muons flux at various depth as well as angular distributions to understand performance of the detector and provide useful inputs to enhance sensitivity for neutrino detection\cite{icecube_muon} \cite{km3net_muon}.

TRIDENT is a next-generation neutrino telescope proposed to be built in the northeastern region of the South China Sea, featuring an innovative detector design and a large sensitive volume that will provide high sensitivity for observing astrophysical neutrino sources and identifying neutrinos of all flavors, with a suitable site identified by the TRIDENT Explorer (T-REX) program in 2021\cite{trident_na_paper}. During the design phase of TRIDENT, understanding the atmospheric muon properties at relevant site will provide valuable inputs for refining the simulation pipeline, improving reconstruction algorithms, and optimizing trigger designs to effectively distinguish neutrino signals from muon backgrounds. As a basis towards this goal, we performed atmospheric muon flux measurements as a function of depth in the sea during the T-REX 2024 program, which a series of sea trials aimed at measuring in-situ optical properties, radioactivity and atmospheric muon background, as well as validating deep-sea techniques such as acoustic and photoelectric transmission. Muon flux measurements at a wide range of sea water depth starting from sea level enable the validation of atmospheric muon generation and propagation in TRIDENT simulation pipeline using real data, it also allows a better estimate of data transmission throughput when designing the trigger strategy. Once upgraded to a device with better angular resolution, which will be explored in future work, MuonSLab can also guide the developments of TRIDENT’s reconstruction method and calibration strategy. As noted in a previous study~\cite{SIMHONY2024169955}, a multi-layer scintillator coupled with sparsely arranged SiPMs can achieve millimeter-level spatial resolution for particle tracking. While muon direction reconstruction is not the primary focus of this work, a detailed investigation into MuonSLab's potential for this application will be explored in future work. Additionally, upgrades to a more pixelated detector configuration are also planned to enhance its directionality reconstruction capabilities.

In this paper, we present an innovative instrument designed for atmospheric muon flux measurement, called MuonSLab, along with its performance at various locations. To facilitate muon measurements in the deep sea and at all altitudes, MuonSLab features a self-contained design and a compact structure for enhanced portability. Tests were conducted to ensure the pressure resistance and water repellency of the deep-sea detector. Given our interest in measuring muon flux from sea level to the ocean floor, and considering that PMTs are not suitable for operation near the surface due to unavoidable environmental light \cite{dumand}, we chose to use four plastic scintillator slabs coupled with silicon photomultipliers (SiPMs) for this application. When a muon passes through the plastic scintillator, it excites scintillation light, which is read out when it reaches the SiPM and causes over-threshold signal. We validated the multi-layer coincidence trigger is sensitive to muon events.

This paper is structured as follows: Section \ref{sec:detector} provides a detailed description of the detector, including sensors, electronics, power supply, data acquisition system, and support structure, while Section \ref{sec:before} presents a series of tests to validate our detectors before performing measurements at sea. Measurements from the sea trials are discussed in Section \ref{sec:sea}, and finally, Section \ref{sec:conclusion} presents the conclusion.

\section{Detector Design}
\label{sec:detector}

\subsection{Mechanics}
\begin{figure}[ht]
    \centering
    \begin{minipage}[b]{0.45\textwidth}
        \centering
        \includegraphics[width=\linewidth]{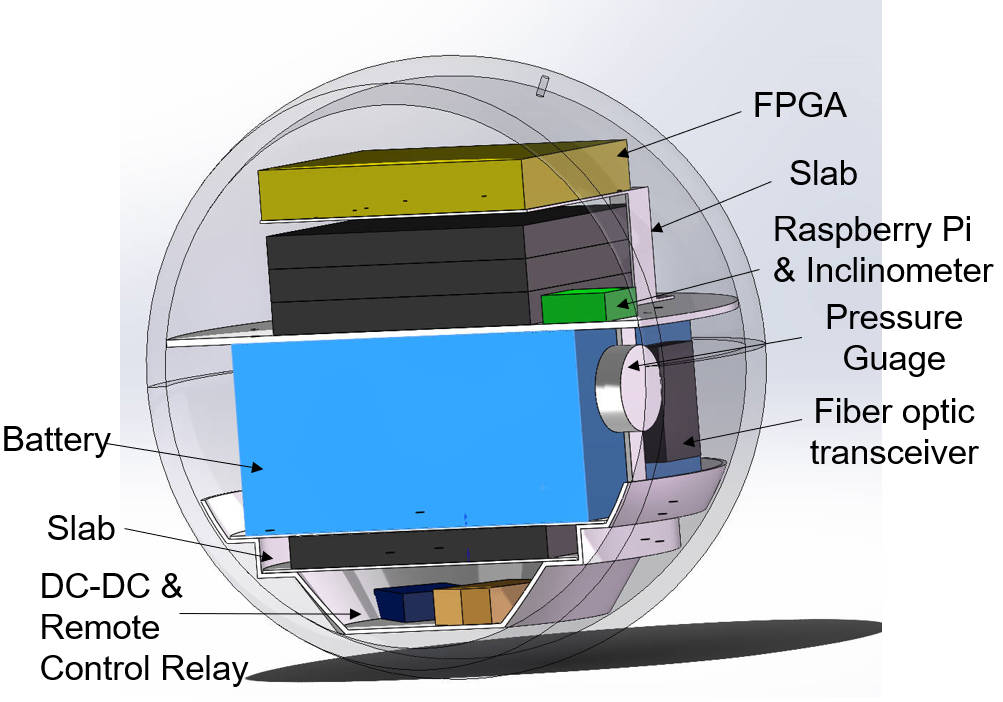}
        \caption{The glass vessel houses the core components of MuonSLab: four plastic scintillator slabs, each couples with eight SiPMs. This self-contained system integrates a battery for power and a data acquisition setup comprising front-end electronics, an FPGA board (for digitization and trigger logic), and a Raspberry Pi.}
        \label{fig:ball}
    \end{minipage}
    \hfill
    \begin{minipage}[b]{0.45\textwidth}
        \centering
        \includegraphics[width=\linewidth]{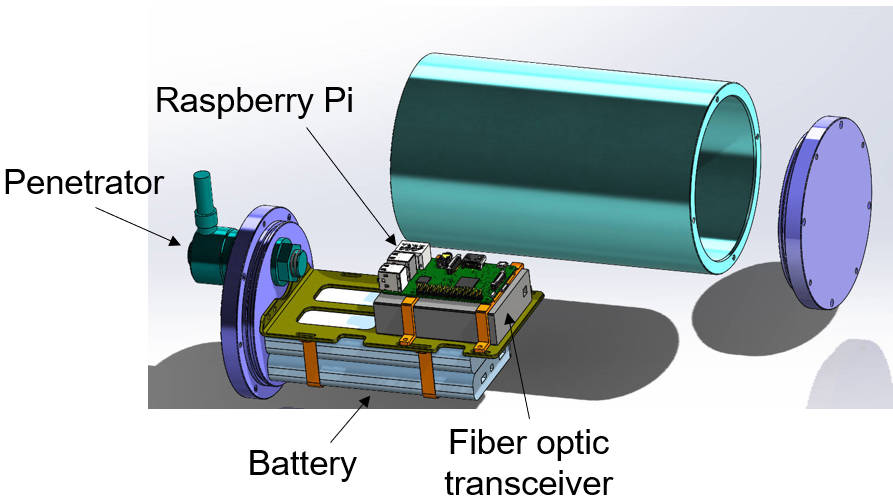}
        \caption{The junction box is connected to the glass vessel via an oil-filled cable housing two electrical wires and an optical fiber. A Raspberry Pi facilitates data backup, while the battery enclosed within the vessel can be recharged via the penetrator system without requiring the vessel to be opened.}
        \label{fig:box}
    \end{minipage}
\end{figure}

All the components are integrated into a spherical glass vessel and a titanium alloy junction box, connected by an oil-filled cable. A sectional view is presented in Figure \ref{fig:ball} and Figure \ref{fig:box}, and detailed information on its structure is provided in the following subsections. The sensors, electronics and power supply devices are enclosed within a 17-inch Vitrovex$^\circledR$ glass sphere produced by Nautilus \cite{ball}, with a diameter of 432 mm and a thickness of 14 mm, which can withstand water pressure up to depths of 6700 m. The glass vessel consists of two hemispheres, with one hemisphere featuring two apertures for the vacuum port and the penetrator. A vacuum of 0.2 atmospheres was created inside the sphere, using the pressure difference to tightly bond the two hemispheres together, and a mechanical pressure gauge is used to monitor the sphere's seal integrity. Data backup instruments are placed in a titanium alloy cylinder junction box, with a diameter of 140 mm and a thickness of 10 mm. An oil-filled cable containing two wires and one optical fiber connects the glass vessel to the junction box via two penetrators. The whole system passed a 40 MPa high-pressure test, as described in Section \ref{subsec:stress test}. Inside the sphere and junction box, we mainly use aluminum alloy materials to ensure strength, save space, and improve heat dissipation.

It is worth noticing that glass containers are essential for deep-sea neutrino telescopes like TRIDENT to house PMTs for Cherenkov light detection but are less suitable for our scintillator detector. Our design choices were driven by the need to validate key engineering components for TRIDENT to assess their feasibility and reliability for future deployment.

\subsection{Sensors}
 There are three sensor systems in MuonSLab: four plastic scintillator slabs targeting atmospheric muons, an inclinometer (JY901S \cite{angle}) measuring the tilt angles of detector, and a depth gauge (RBR95950 \cite{depth}) providing water depth measurement. For the most important part, plastic scintillator slabs, we chose the SP101 plastic scintillator from Beijing Hoton Nuclear Technology, Co., Ltd \cite{SP101}. It's based on polystyrene with a peak emission wavelength of 423 nm and a light output of 64\% (Anthracene). To maximize the detection area within the limited sphere space, we made each slab to the size of 200 mm $\times$ 200 mm $\times$ 20 mm, and placed three slabs above the battery and one underneath to utilize triple-layer coincidence readout for the suppression of signals from low-energy electrons or gammas. Our detector deploys the On Semiconductor MicroFC 60035 C-Series 6 mm $\times$ 6 mm SiPM \cite{SiPM}, which features a gain of $3 \times 10^6$, low noise and good temperature stability. Using an oscilloscope, we determined that the amplitude of single photon events (SPE) is 0.4 mV. Each slab has eight SiPMs evenly distributed on its narrow side face, and the signals from four SiPMs on adjacent sides are amplified 40 times by the electronic board and then directly summed to produce one analog input to an FPGA board. 

\subsection{Electronics}
MuonSLab relies on three types of electronic boards to operate: eight customized front-end amplification and summation boards coupled with 32 SiPMs for muon signal identification, an ALINX AXKU040 FPGA development board \cite{FPGA} connected with two ALINX FL9627 ADC boards \cite{ADC} for data acquisition and two Raspberry Pi 4B units for the inclinometer and data backup. 

\begin{figure}
    \centering
    \includegraphics[width=\linewidth]{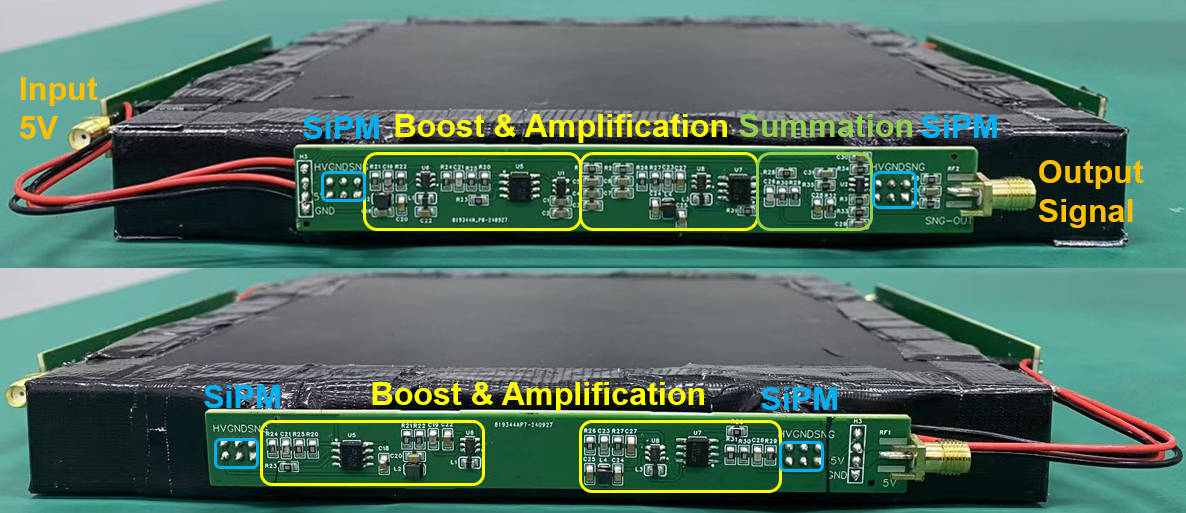}
    \caption{Front-end Electronics for MuonSLab, 8 sets are used in the detector, each of them is taking care of powering 4 SiPMs on the adjacent edges in one layer, and subsequent amplification of the analog signals, which are then summed before sending out for digitization.}
    \label{fig:electronics}
\end{figure}

A front-end electronics board consists of five functional modules. Four of them are identical boost and amplification modules, which step up the input 5V voltage to 29.5V to power the SiPMs and amplify the SiPM signals by a factor of 40, inspired by the CosmicWatch program \cite{CosmicWatch}. The fifth module is a reverse amplifier circuit with a gain of one, which directly sums the four amplified SiPM signals to serve as the input for the ADC board (4-in-1 configuration). Fabricated as two 16 mm $\times$ 140 mm strips connected by four wires, the board can be fixed around the slab with pin headers, which greatly saves space and simplifies assembly and debugging, as shown in Figure \ref{fig:electronics}.

The waveforms of SiPMs are read at a sampling rate of 125 MHz using the 8-channel ADC board in combination with the ALINX FPGA board. A hardware trigger is implemented in the FPGA, requiring at least three slabs to have over-threshold signals. The trigger strategy was designed to minimize data transmission bandwidth on the premise of not losing the signals of interest, considering that muons can traverse three stacked slabs, whereas other charged particles, such as electrons, cannot. The threshold here was a certain voltage ensuring that most dark noise would remain below this level. After triggering, the waveforms of all 8 channels are packaged as an event and transmitted to a micro SD card and two Raspberry Pis, with one of the Raspberry Pi in the glass vessel and the other one in the junction box for data backup.

\begin{figure}[H]
    \centering
    \includegraphics[width=\linewidth]{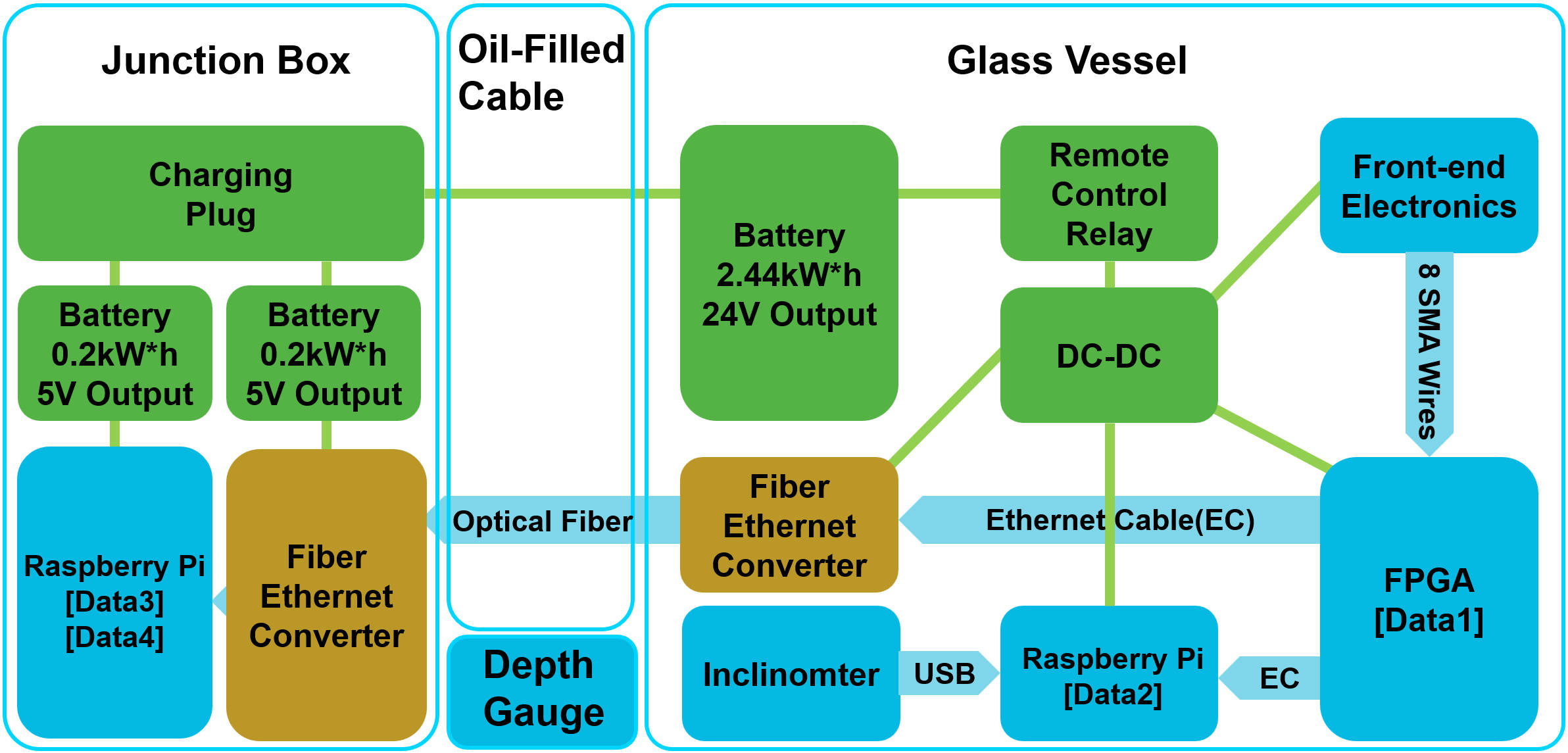}
    \caption{Power supply and data flow of MuonSLab. The green lines represent electrical cables among various components and the blue arrows show data flow from front-end electronics to Rasberry Pi that takes care of final data storage and backup.}
    \label{fig:PowerAndData}
\end{figure}

The need for rapid deployment and high reliability guided our choice of a self-contained design, utilizing batteries as the power source. In the glass vessel, we chose a 270 mm $\times$ 194 mm $\times$ 133 mm Li-ion battery with a capacity of 2.44 kW·h, which has Bluetooth functionality, allowing us to monitor the remaining battery level and discharge status using a mobile phone. A dual-channel remote control relay is used to control the circuit inside the sphere via a remote controller and we connected the two channels of the remote control relay in parallel for redundancy to reduce the risk of relay failure. 
Inside the junction box we use two 0.2 kW·h small Li-ion batteries to power a Raspberry Pi and a fiber-optical transceiver separately. There's also a charging plug in the junction box connected to the battery inside the glass sphere via the two wires inside the oil-filled cable, allowing us to charge the batteries by opening the junction box even after sealing the sphere. The total power consumption is 21 W inside the sphere and 7 W in the junction box, meaning they can operate for about 5 days and 2 days continuously, which is sufficient for our needs. A schematic diagram of power supply and data flow of MuonSLab is shown in Figure \ref{fig:PowerAndData}.

\section{Detector Performance}
\label{sec:before}
\subsection{Sensitivity Tests}
\label{subsec:sensitivity}
\begin{figure}[H]
    \centering
     \begin{minipage}{0.4\textwidth}
        \centering
        \includegraphics[width=\linewidth]{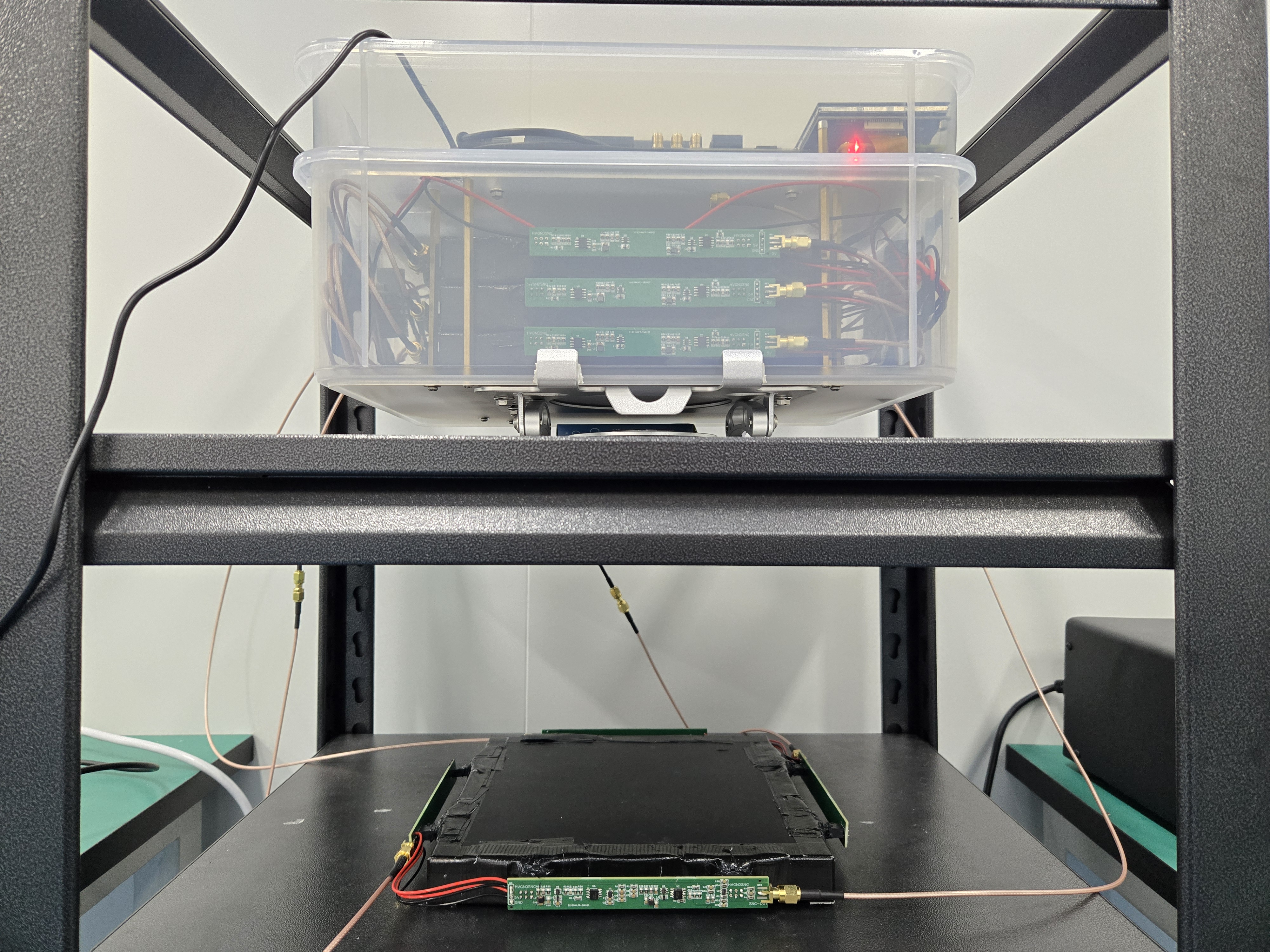}
    \end{minipage}
    \begin{minipage}{0.5\textwidth}
        \centering
        \includegraphics[width=\linewidth]{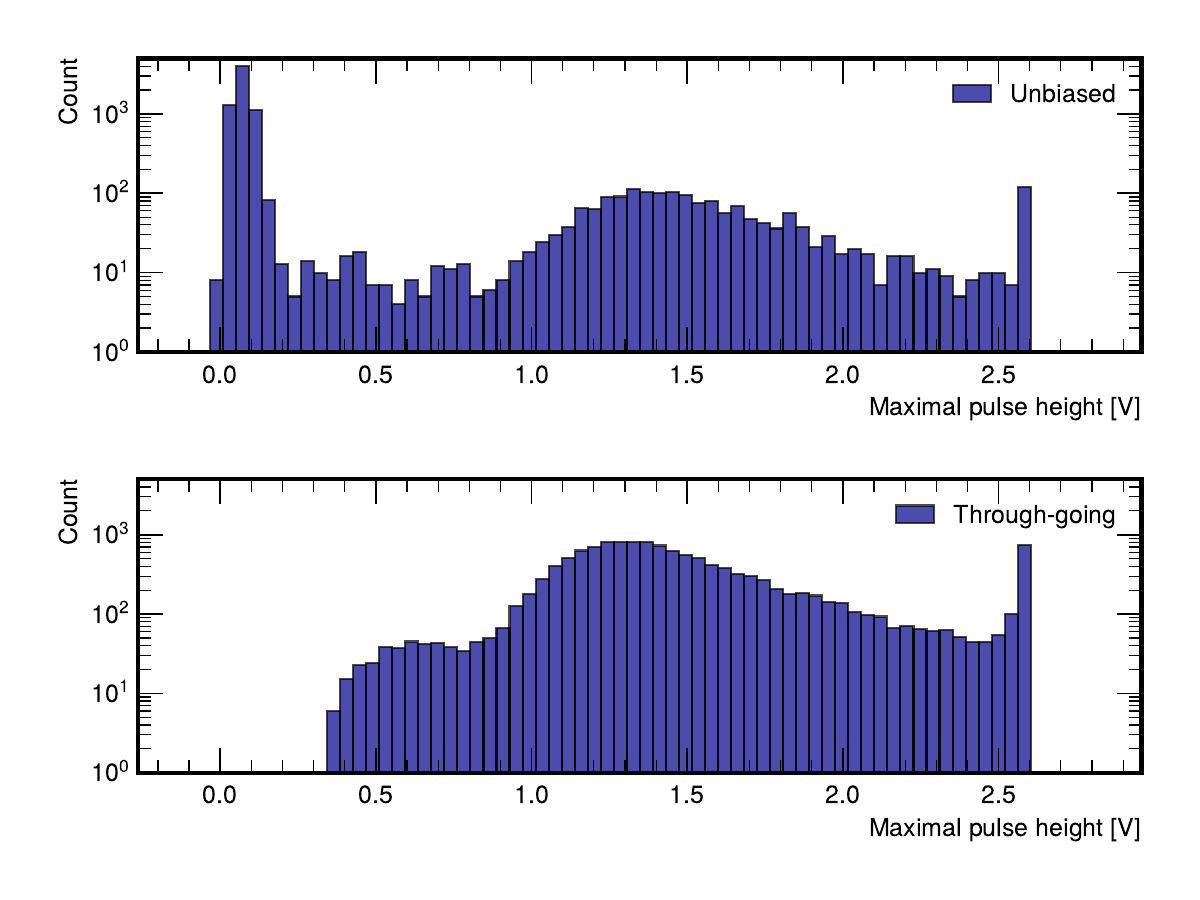}
    \end{minipage}
    \hfill
    \caption{Left: The relative distance between slabs in the glass vessel was repeated on shelf. Upper right: The inclusive amplitude distribution of one channel from the bottom slab including both muon signal and background noise. Lower right: The amplitude distribution of one channel from the bottom slab requiring all channels were over-threshold, indicating through-going muon events.}
    \label{fig:noisepeakmea}
\end{figure}

The flux of atmospheric muons decreases rapidly with sea depth, as shown in both theoretical predictions \cite{muon_theoretical} and former measurements \cite{muon_experimental}. The typical number of induced photon for a minimum ionizing particle muon traversing a thin scintillator follows a Landau distribution, exhibiting a prominent peak known as the Muon Peak in the maximal output voltage distribution from muons. A detector sensitive to muons should be able to distinguish the Muon Peak from the noise distribution in the amplitude distribution. We conducted a validation test in the lab to assess the sensitivity of the 4-in-1 configuration (Figure \ref{fig:noisepeakmea}). 

In this test, we repeated the configuration of 4 slabs but remove the battery and changed the trigger strategy to require that at least two slabs have channels exceeding the threshold. Our analysis concentrated on data in which the middle two slabs exhibited over-threshold channels, while the top slab remained below threshold. This selection indicated that the entire system was triggered by a highly tilted muon, which rarely impacted the bottom slab, thereby providing an inclusive readout results for the bottom slab, including both muon and background noise. The amplitude distribution of one channel from the bottom slab showed that most recorded information was from noise, resulting in a peak below the threshold of 0.37 V. To identify the second peak in that results, we made another selection that all channels were over-threshold, indicative of through-going muon presence and greater than 4 MeV energy deposition in one slab. This analysis revealed a muon peak in the amplitude distribution of one channel from the bottom slab, suggesting that the second peak in the former selection also corresponds to a muon peak, which can be distinctly separated from the noise distribution. Moreover, considering the primary radioactivity background in seawater, particularly from K40, which can induce gamma radiation of 1.46 MeV or electrons up to 1.3 MeV, the most probable output voltage for a single channel assuming all background energy is deposited in one slab is just near the trigger threshold. This indicates that it is unlikely to trigger the three-layer requirement. Thus, these findings demonstrate that the three-layer trigger strategy and threshold selection are effective for detecting muons. Using Geant4~\cite{ALLISON2016186}, we performed simulations modeling the MuonSLab detector configuration and simulating K40 decay at an activity level of 10.78 Bq/kg~\cite{trident_na_paper}, simulation result demonstrated that the K40 induced trigger rate remains well below 0.001 Hz, which is the trigger rate we'd expect for measurement performed at 2000 m below sea level, and the background rejection efficiency is measured to exceed 99.9\%.


\begin{figure}[hb]
    \centering
     \begin{minipage}{0.2\textwidth}
        \centering
        \includegraphics[width=\linewidth]{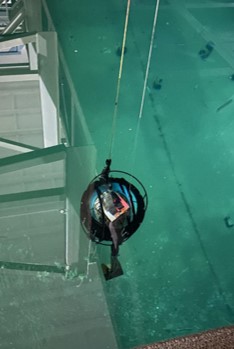}
    \end{minipage}
    \begin{minipage}{0.7\textwidth}
        \centering
        \includegraphics[width=\linewidth]{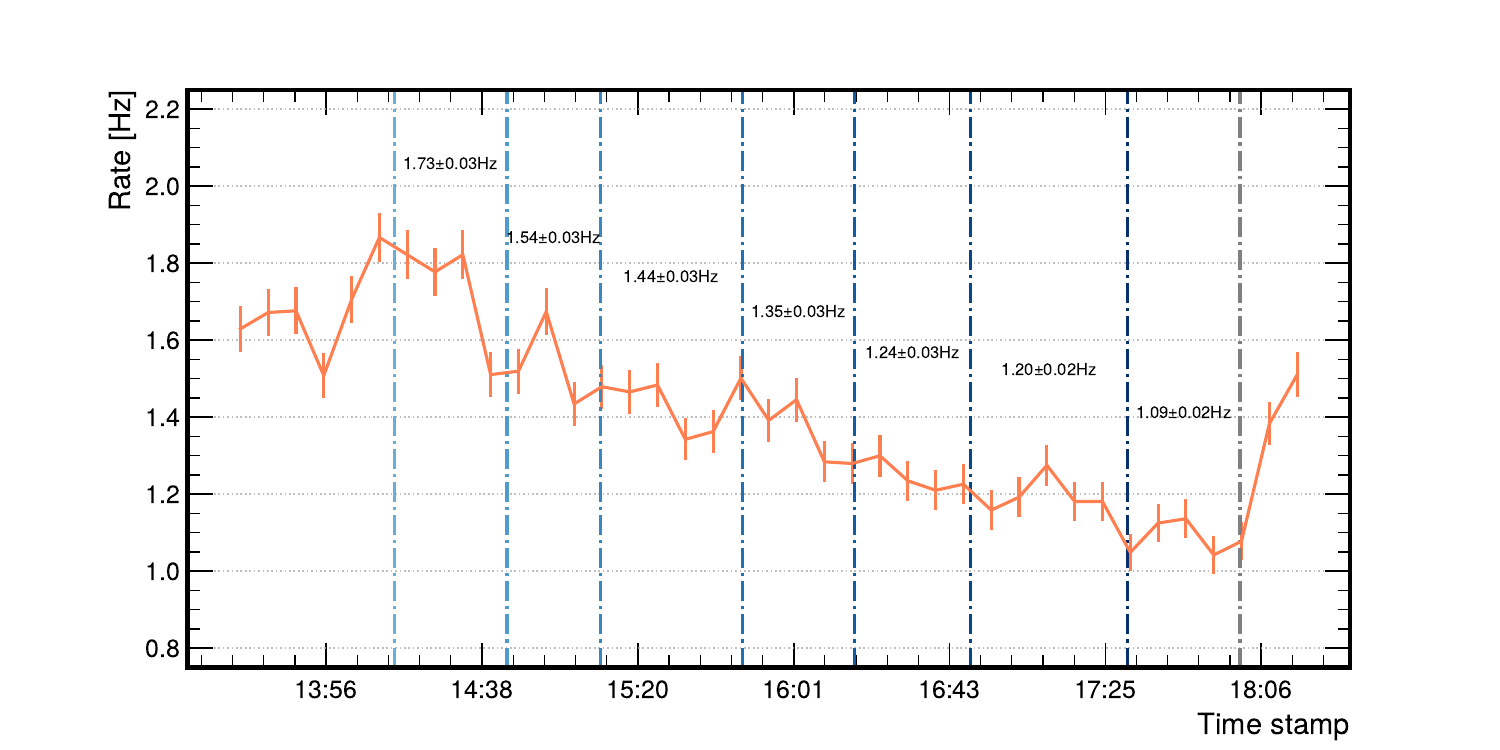}
    \end{minipage}
     \caption{Result of water tank measurement, each interval between two blue dot-dashed lines indicate a certain depth, starting from water surface to 6 meters below the surface, then the MuonSLab was taken up from the bottom of the water tank around 18:00.}
    \label{fig:watertankresult}
\end{figure}

To validate the device is sensitive to muon flux variation with different overburden, MuonSLab as in the final configuration was tested in an 8-meter deep ship towing tank located on the campus of Shanghai Jiao Tong University before its deployment to the deep sea. The results (Figure \ref{fig:watertankresult}) show that muon counts consistently decreased as the detector's depth increased from 1 to 6 meters. This demonstrates that the system operates effectively underwater and is sensitive to changes in muon flux with respect to depth.

\subsection{Stress Tests}
\label{subsec:stress test}
Before deploying the detector into the ocean, we tested its resilience to deep-sea environmental conditions, focusing on the risks associated with high pressure, water leakage, and low temperatures. We conducted a 40 MPa pressure test while the entire detector was operating at the Underwater Engineering Institute on the Shanghai Jiao Tong University campus. The same method was used for both increasing and decreasing pressure, with the pressure changed in 5 MPa and held for 10 minutes each time to thoroughly test the system's performance under different pressures. When the maximum pressure was reached, we maintained it for 20 hours. The result was successful, with no water leakage and MuonSLab worked well during the whole period. According to the measurements from the T-REX experiment, the water temperature in the deep sea is about 2 degrees Celsius. A validation test for the detector was conducted in a cold storage room at temperatures ranging from 2 to 3 degrees Celsius, lasting for more than 48 hours. During this time, the detector continued to operate normally, showing that the electronics and mechanical structures maintain functional stability at the temperature of deep sea.

\section{Measurements}
\label{sec:sea}
This section presents the measurements conducted with MuonSLab during T-REX 2024. MuonSLab was deployed to the northeastern region of the South China Sea. The detector was successfully retrieved from the deep sea. Additionally, auxiliary measurements were conducted using a similar detector placed in a portable box, demonstrating the extended use case of MuonSLab.

\subsection{Sea Trial Measurement}

\begin{figure}[H]
    \centering
    \begin{minipage}{0.6\textwidth}
        \centering
        \includegraphics[width=0.8\linewidth]{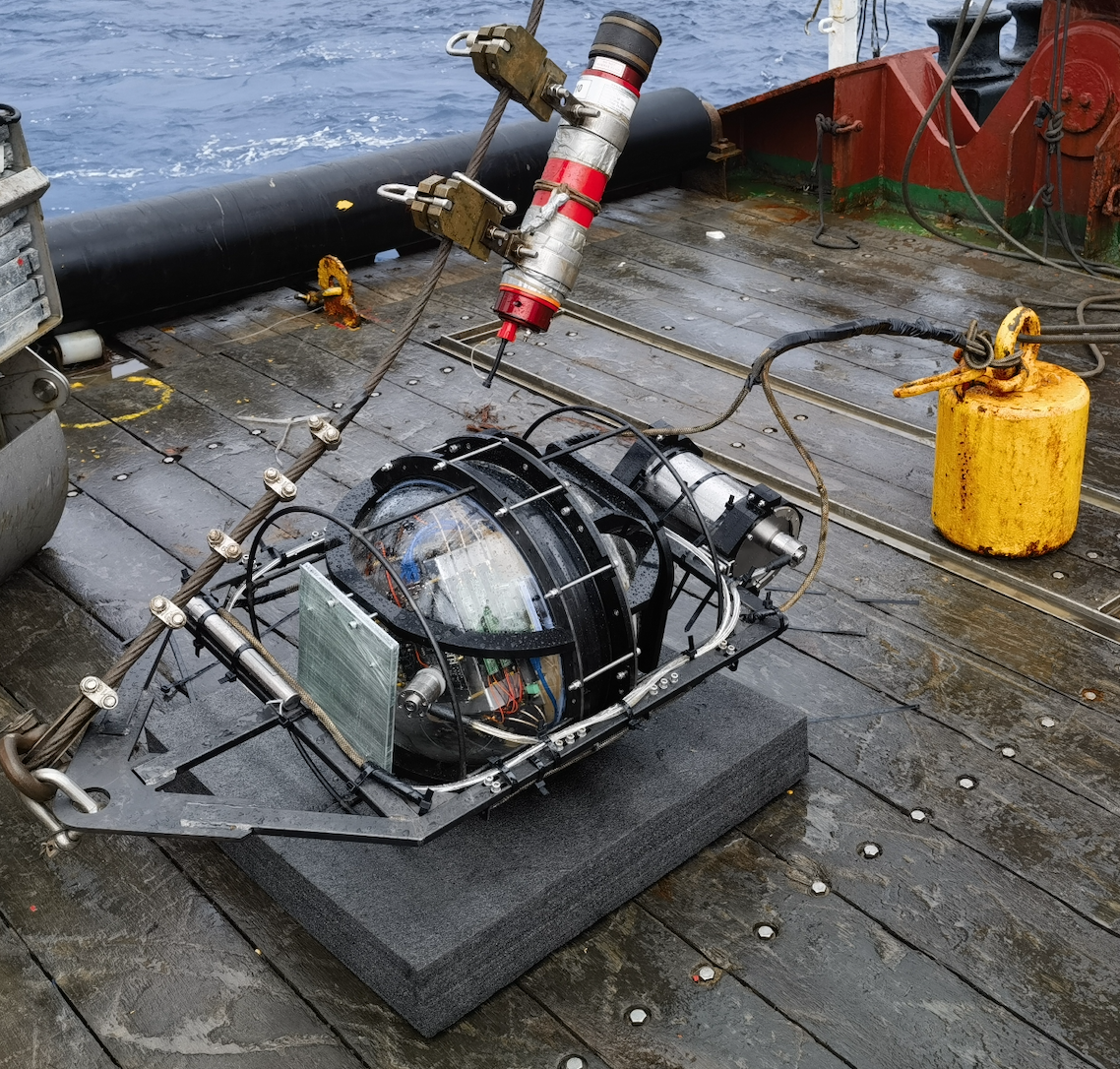}
    \end{minipage}
    \begin{minipage}{0.3\textwidth}
        \centering
        \includegraphics[width=0.7\linewidth]{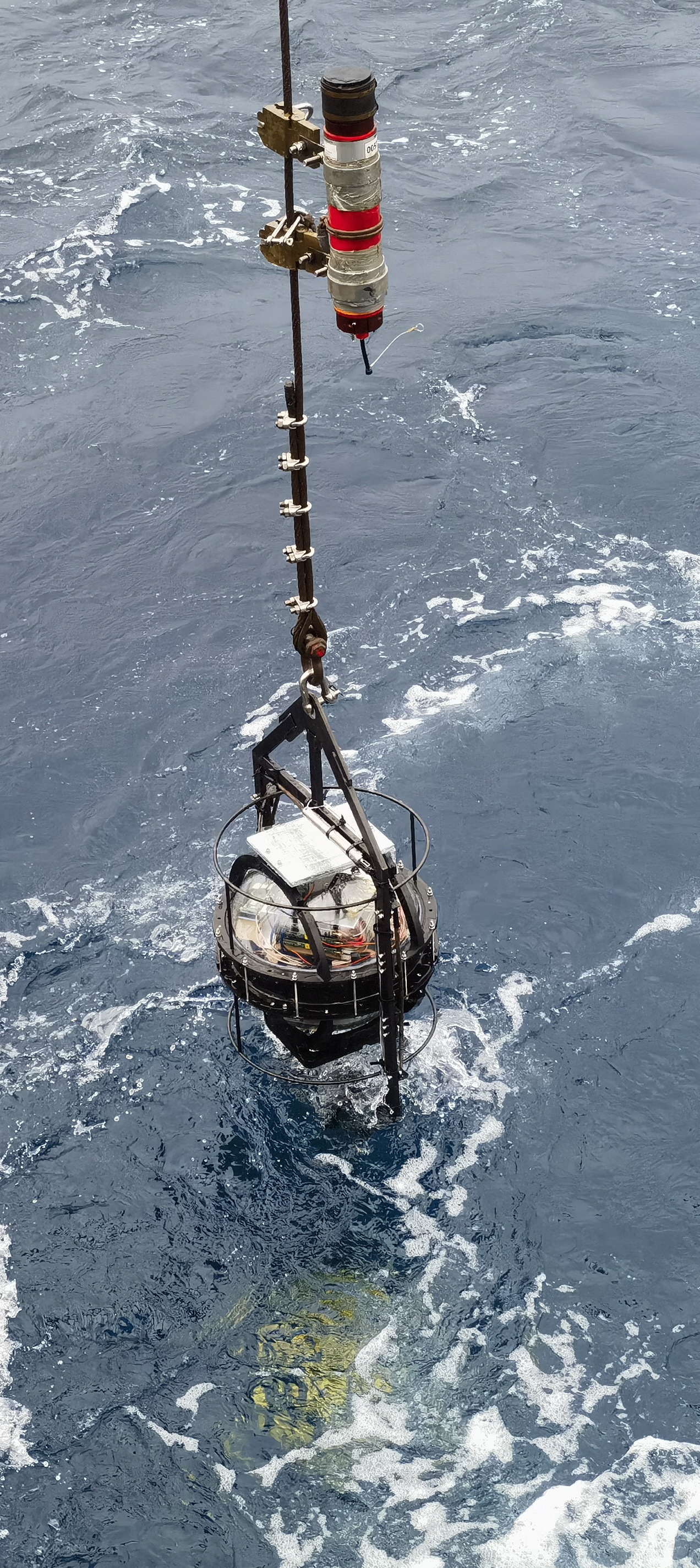}
    \end{minipage}
    \hfill
    \caption{The pressure-resistant vessel containing the device was placed on the deck before deployment (left) and then lowered into the sea water (right).}
    \label{fig:into_water}
\end{figure}

\begin{figure}[H]
    \centering
    \begin{minipage}{0.49\textwidth}
        \centering
        \includegraphics[width=0.9\linewidth]{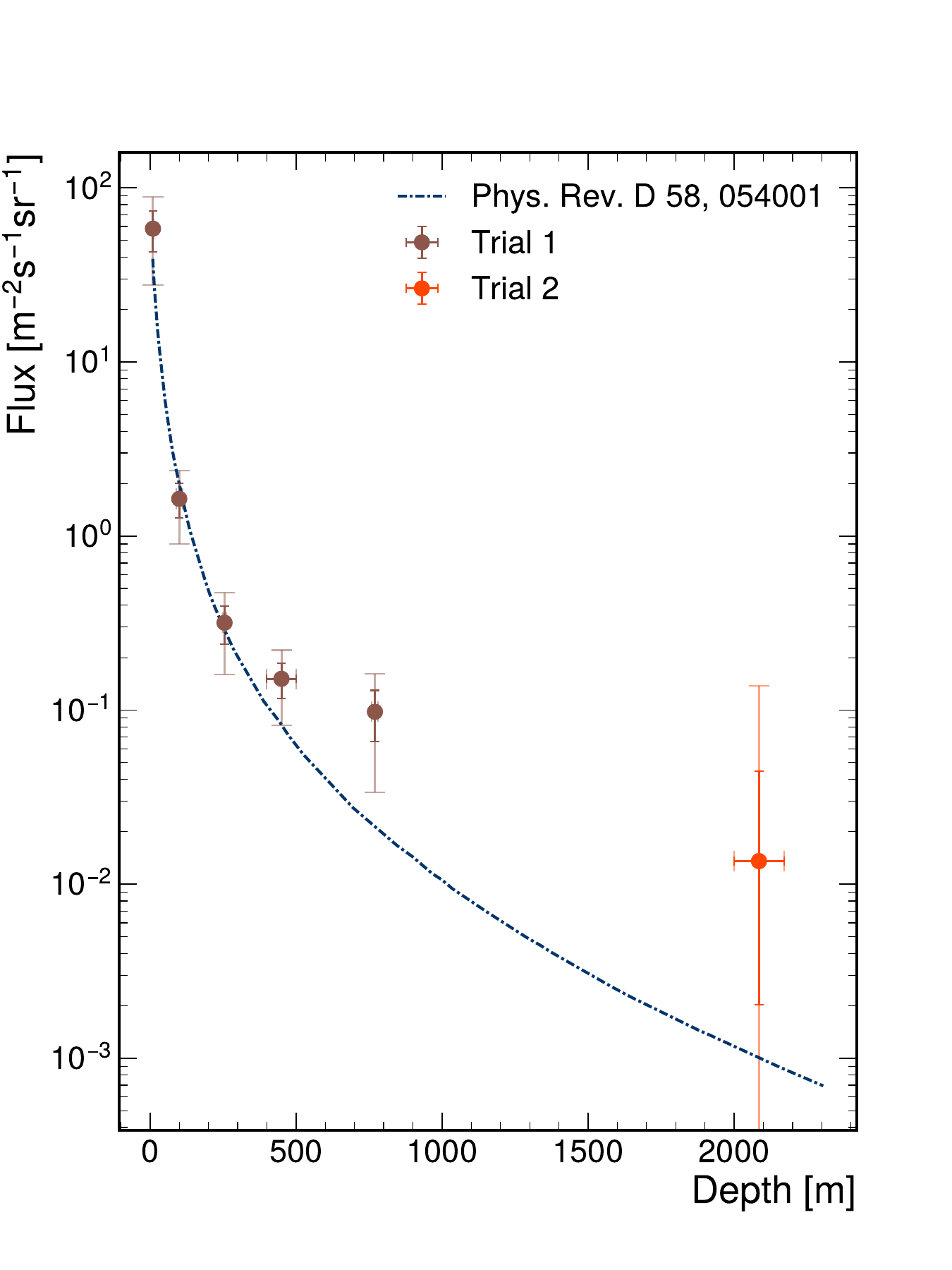}
    \end{minipage}
    \begin{minipage}{0.49\textwidth}
         \centering
         \includegraphics[width=0.95\linewidth]{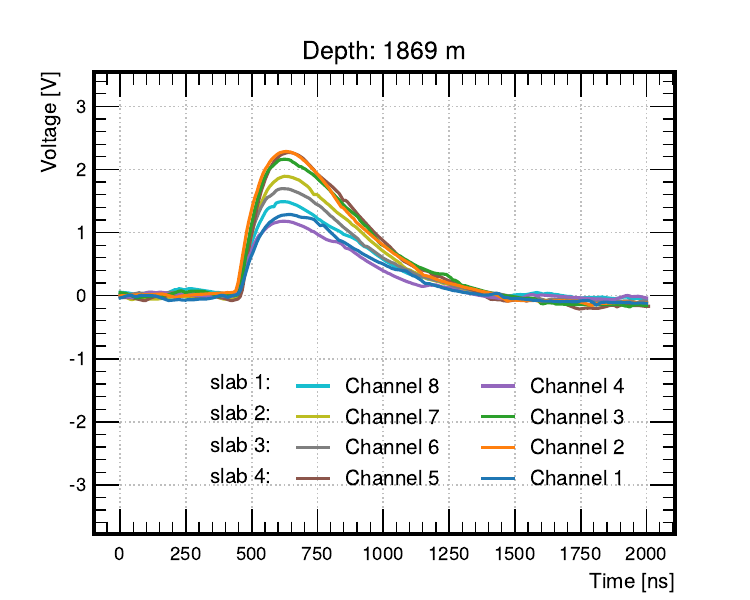}
    \end{minipage}
    \hfill
    \caption{Left: Muon flux as a function of depth is presented using data from the two sea trials. In the first trial, the detector was paused at several depths, while in the second trial, it was paused only at approximately 2.1 kilometers. Reference flux from previous experiments~\cite{PhysRevD.58.054001} has been included for comparison. Vertical error bars represent 1 $\sigma$ (darker) and 2 $\sigma$ (lighter) total uncertainty, with both statistical and systematic uncertainty considered. Measurements at around 480 and 2200 meters have larger horizontal error bar due to the device's non-negligible vertical movements in these depths while in the sea. Right: An event display of a muon candidate at depth of 1869 m when the detector was being lowered. The accompanying legend indicates the relative positions of the channels within each slab, with "slab 1" corresponding to the top layer and "slab 4" being the bottom layer beneath the battery. The fact that all channels were over-threshold indicates that a muon passed through all four slabs.}
    \label{fig:moneyplot}
\end{figure}

During T-REX 2024, two deployments were made for MuonSLab. The pressure-resistant vessel was attached to a steel cable, allowing it to be lowered into the water at a uniform speed, pausing at specified depths to collect data (Figure \ref{fig:into_water}). The first trial lasted approximately four hours, during which more than 1,500 muons were recorded, while the second trial lasted about two hours and collected 357 muons. During the first trial, the MuonSLab was held at various specific depths for a designated duration, ranging from sea level to 800 meters. The second sea trial was conducted at the Hai-Ling Basin~\cite{trident_na_paper}, where the same detector was dropped directly to a depth of 2 km and remained at that depth for a period of time. The muon identification is based on a selection criterion that requires any three layers of slab to exhibit over-threshold channels. Despite the low muon flux in the deep sea, we performed muon counting measurements at some specified depths all the way to 2.1 km below sea level, which gives us the muon counting rate varying with depth (Figure \ref{fig:moneyplot}).

The flux measurement from this sea trial is still dominated by statistical uncertainty, especially at deep locations. The main systematic uncertainty comes from counting inefficiency due to the choice of threshold and solid angle that MuonSLab could cover during the sea trial, which were estimated to be 11\% and 20\% respectively. The inefficiency per layer is calculated by requiring one layer not having over-threshold pulse while the other three layers have. While the uncertainty due to solid angle comes from a distribution it that covers all possible angles that a muon could traverse the MuonSLab.

The measurements performed during the two sea trials last only a few hours in total. To ensure the long-term operation of MuonSLab, additional mitigation strategies will need to be implemented. While the stainless steel supports used in these trials showed no signs of corrosion during their brief deployment, titanium alloy—a more corrosion-resistant material—will replace stainless steel components in future applications. Prior to deployment, prolonged corrosion testing will also be conducted. On the other hand, since MuonSLab employs a scintillator-based detection method instead of relying on detecting Cherenkov light in seawater, biofouling poses minimal risk to its operation. To further ensure reliability, dedicated calibration devices (e.g., an LED-based system for monitoring SiPM response) will be integrated for real-time performance tracking. These systems will also provide data for subsequent offline corrections to measured observables.

\subsection{Auxiliary Measurements}
\begin{figure}[H]
    \centering
    \begin{minipage}{0.45\textwidth}
        \centering
        \begin{minipage}{\linewidth}
            \centering
            \includegraphics[width=0.8\linewidth]{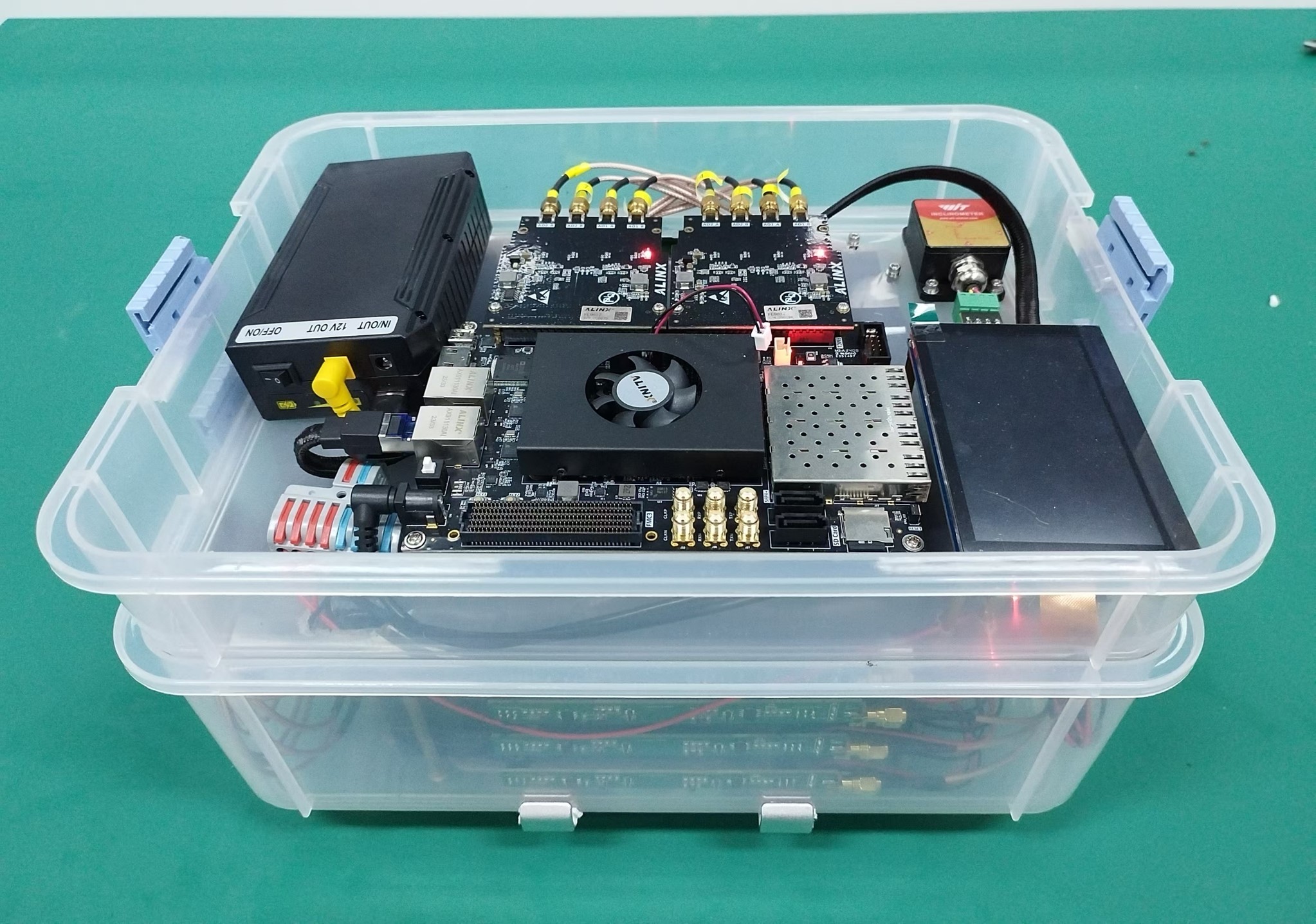}
            \label{fig:left_top}
        \end{minipage}
        \vspace{1cm} 
        \begin{minipage}{\linewidth}
            \centering
            \includegraphics[width=0.8\linewidth]{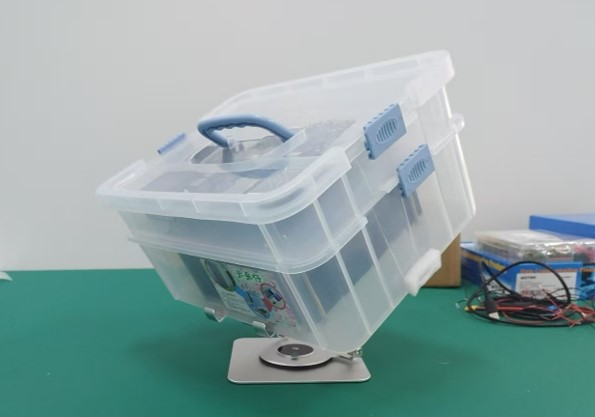}
            \label{fig:left_bottom}
        \end{minipage}
    \end{minipage}
    \hfill
    \begin{minipage}{0.5\textwidth}
        \begin{minipage}{\linewidth}
        \centering
        \includegraphics[width=1.1\linewidth]{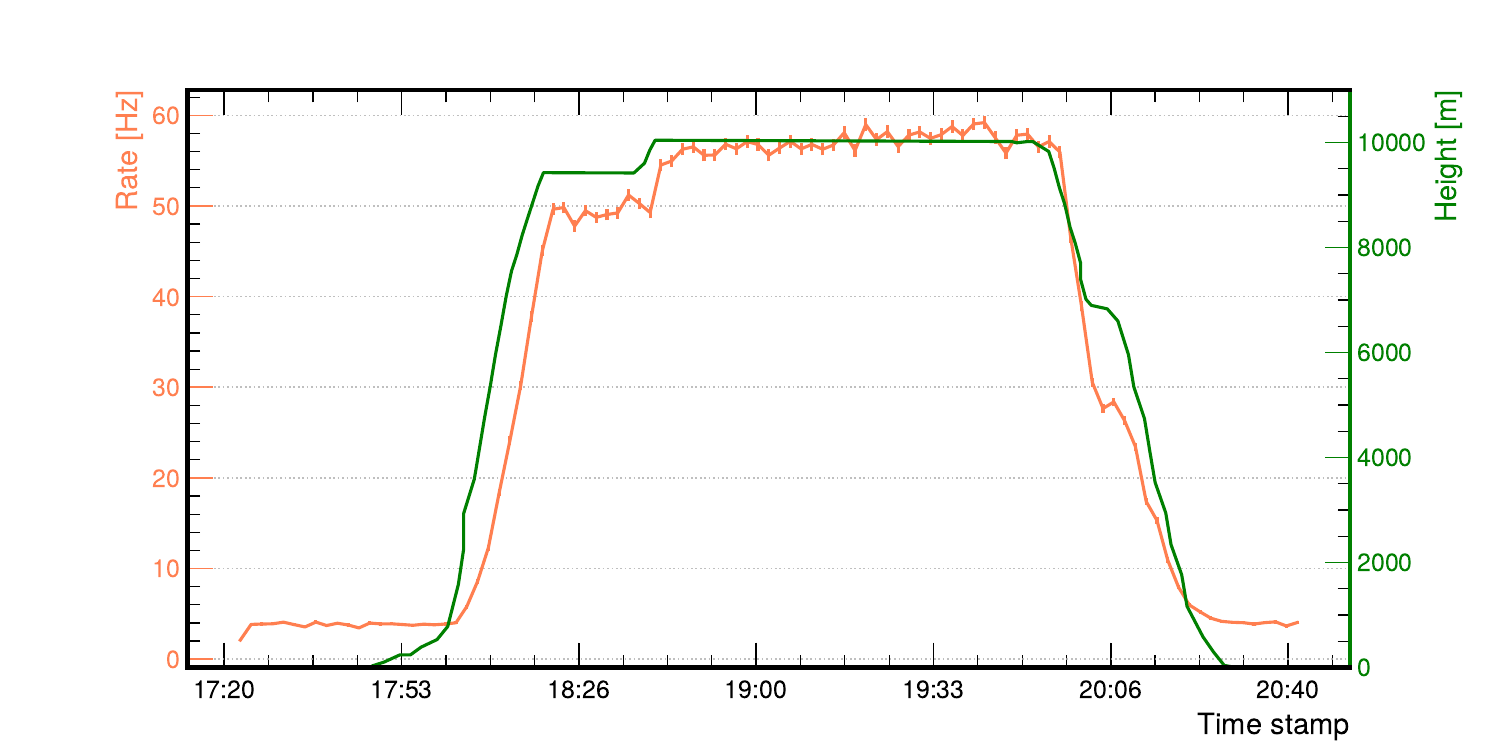}
        \label{fig:right}
        \end{minipage}
        \hspace{0.1cm} 
        \begin{minipage}{\linewidth}
            \centering
            \includegraphics[width=1.1\linewidth]{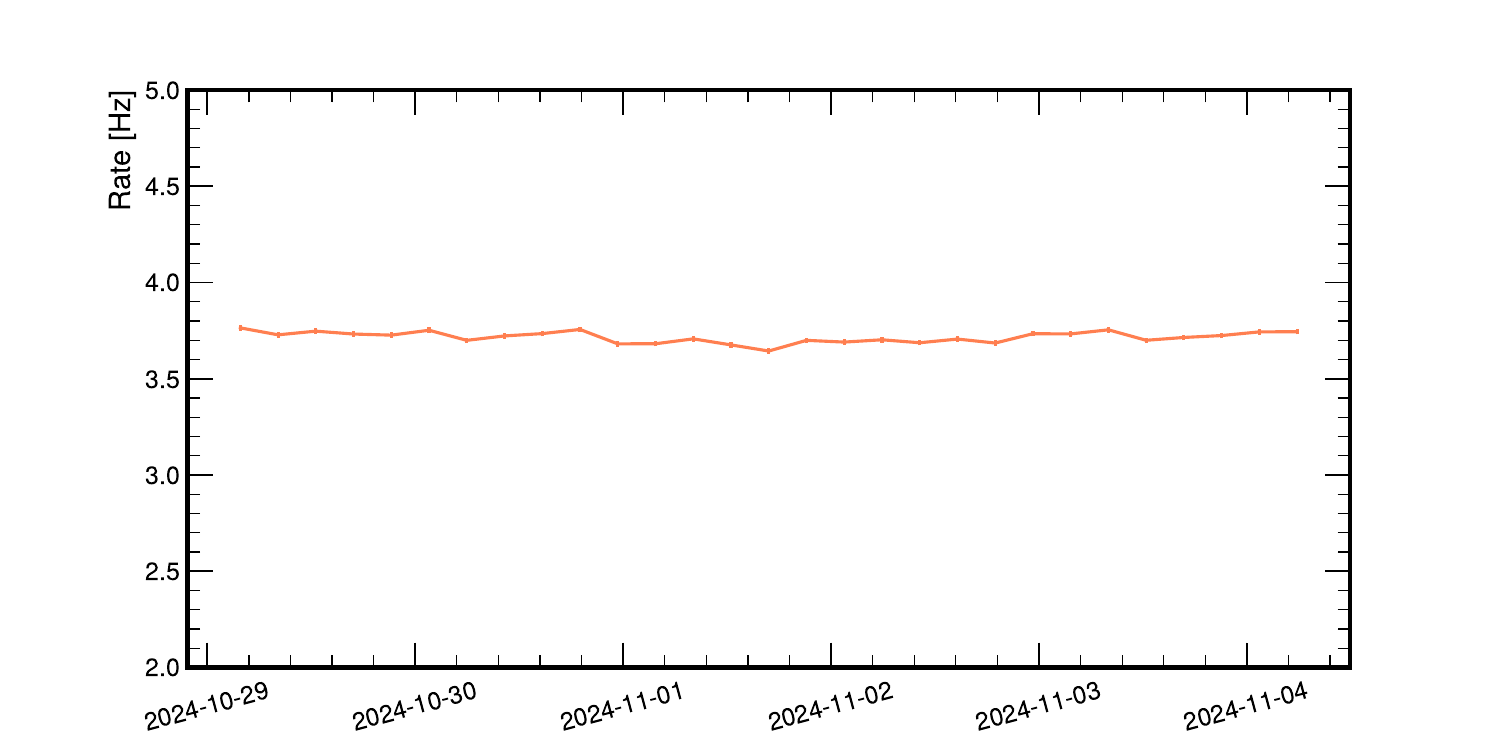}
            \label{fig:left_bottom}
        \end{minipage}
    \end{minipage}
    \caption{Left: Demonstration of the MuonSLab housed in a portable case. Upper right: Muon counting rate (orange) during a flight, accompanied by height variation (green). Lower right: Muon counting rate over a week of continuous operation.}
    \label{fig:combined}
\end{figure}

A set of MuonSLab, along with an inclinometer, was integrated into a portable case, maintaining its functionalities for muon counting and inclination measurement. We simplified the support structure and reduced the battery size, transforming the deep-sea device into a portable unit that can be taken on a plane from Sanya to Shanghai. This allows for measuring muon variations in counting rate during flight, from ground level to an altitude of 10,000 meters. During T-REX 2024, this device was placed in the cockpit and collected data for about a week, yielding stable results. These measurements demonstrate the ability of our detectors to operate stably over long periods of time, as well as the flexibility to measure muon in a variety of environments. (Figure \ref{fig:combined})

\section{Conclusion}
\label{sec:conclusion}
MuonSLab, as a compact muon detector, has demonstrated its sensitivity and stability in measuring muons. The successful deployment of this device during T-REX 2024 confirmed our ability to accurately measure muon flux at various depths in the South China Sea. The initial experimental results align well with our expectations, and we are currently conducting simulations and further analysis to derive more physical quantities related to depth. The portable version of our detector, which exhibited remarkable stability during T-REX 2024, will play a crucial role in future exploration of using plastic scintillator in the deep sea experiment to understand muon properties associated with cosmic rays and extensive air showers. MuonSLab can also provide valuable input to the developments of TRIDENT’s simulation and trigger systems. Furthermore, the detector facilitates the use of plastic scintillators for particle identification in the deep sea, offering preliminary experience for future TRIDENT calibration schemes and physics potential exploration.

\appendix

\acknowledgments

This work is supported by the National Natural Science Foundation of China (No. 12427809) and the National key research and development plan (No. 2023YFC3107401). The authors thank Jun Guo and Kim Siang Khaw for their help in improving the manuscript. The authors appreciate crew from Haiyang Dizhi Sihao research vessel for offshore deployment and are also grateful to the Underwater Engineering Institute and the School of Ocean and Civil Engineering of Shanghai Jiao Tong University for providing the pressure testing platform and the water tank. The authors thanks Beijing Hoton Nuclear Technology, Co., Ltd for organic scintillators production and packaging,  Epic Crystal for inorganic scintillators, the Second Institute of Oceanography at Ministry of Natural Resources for the depth gauge.

\bibliographystyle{JHEP.bst}
\bibliography{biblio}

\end{document}